\documentclass[twocolumn,showpacs,preprintnumbers,amsmath,amssymb,epsf]{revtex4-1}

\usepackage{graphicx,epsfig}
\usepackage{dcolumn}
\usepackage{bm}
\newcommand{\be}{\begin{equation}}
\newcommand{\ee}{\end{equation}}
\newcommand{\bse}{\begin{subequations}}
\newcommand{\ese}{\end{subequations}}
\newcommand{\bary}{\begin{eqnarray}}
\newcommand{\eary}{\end{eqnarray}}
\newcommand{\bwt}{\begin{widetext}}
\newcommand{\ewt}{\end{widetext}}

\begin{document}


\title{Model-dependent high-energy neutrino flux from Gamma-Ray Bursts}
\author{Bing Zhang$^{1,2,3}$ and Pawan Kumar$^{4}$}
\affiliation{
$^{1}$Kavli Institute for Astronomy and Astrophysics,
Peking University, Beijing 100871, China\\
$^{2}$Department of Astronomy, School of Physics, 
Peking University, Beijing 100871, China\\
$^{3}$Department of Physics and Astronomy, University of Nevada, Las Vegas, NV 89154, USA, zhang@physics.unlv.edu\\
$^{4}$Department of Astronomy, University of Texas at Austin, TX 78712, USA, pk@surya.as.utexas.edu}


\begin{abstract}
The IceCube Collaboration recently reported a stringent upper limit on
the high energy neutrino flux from GRBs, which provides a meaningful
constraint on the standard internal shock model. Recent broad band
electromagnetic observations of GRBs also challenge the internal shock
paradigm for GRBs, and some competing models for $\gamma$-ray prompt
emission have been proposed.  We describe a general scheme for
calculating the GRB neutrino flux, and compare the predicted neutrino
flux levels for different models.  We point out that the current
neutrino flux upper limit starts to constrain the standard internal
shock model. The dissipative photosphere models are also challenged if
the cosmic ray luminosity from GRBs is at least 10 times larger than
the $\gamma$-ray luminosity. If the neutrino flux upper limit
continues to go down in the next few years, then it would suggest the
following possibilities: 1. the photon-to-proton luminosity ratio in
GRBs is anomalously high for shocks, which may be achieved in some
dissipative photosphere models and magnetic dissipation models; or
2. the GRB emission site is at a larger radius than the internal shock
radius, as expected in some magnetic dissipation models such as the
ICMART model.
\end{abstract}

\maketitle

{\em I. Introduction. ---}
As energetic, non-thermal photon emitters, gamma-ray bursts (GRBs)
have long been regarded as efficient cosmic ray accelerators 
\cite{waxman95}. Assuming that protons and photons are roughly
isotropic in the co-moving frame, significant neutrino emission 
is possible via the $p\gamma$ mechanism at the $\Delta$-resonance,
if protons in a GRB jet can be accelerated to
an energy $E_p$ to satisfy the condition
\begin{equation}
E_p E_\gamma \sim \frac{m_\Delta^2 - m_p^2}{2} \left(\frac{\Gamma}
{1+z}\right)^2 = 0.147 ~{\rm GeV}^2 \left(\frac{\Gamma}{1+z}\right)^2.
\label{pgamma}
\end{equation}
Here $\Gamma$ is the bulk Lorentz 
factor, $E_\gamma$ and $E_p$ are photon and proton energies in 
the observer frame, $m_\Delta=1.232$ GeV 
and $m_p = 0.938$ GeV are the rest masses of $\Delta^+$ and proton, 
respectively. 
For GRBs, a guaranteed target photon source for $p\gamma$
interaction is the burst itself. For the 
typical peak photon energy $E_\gamma \sim$ several hundred keV, the 
corresponding neutrino energy 
\begin{equation}
E_\nu \simeq 0.05 E_p
\label{EnuEp}
\end{equation} 
is in the sub-PeV regime \cite{waxman97,murase} which is well suited 
for detection with the current high-energy neutrino detectors such as 
the IceCube \cite{ahrens04}.
Indeed over the years, the IceCube Collaboration have been searching
for high energy neutrino signals coincident with GRBs in time and
direction, and progressively deeper non-detection upper limits have
been placed \cite{icecube1,icecube2}, which are now beginning to constrain the 
standard GRB internal shock model \cite{waxman97}. The current IceCube 
upper limit was claimed to be at least a factor of 3.7 smaller 
than the theoretical predictions 
for the neutrino flux from GRBs according to the internal-shock model
if the proton luminosity in the shock is normalized to
allow GRBs to account for the flux of UHECRs. The upper limit 
therefore casted a doubt regarding the viability
of GRBs as the main source of UHECRs \cite{icecube2}.
More detailed, follow-up, calculations \cite{li12,hummer12,he12} 
suggest that the current limit is still not deep enough to provide 
significant constraints on the validity of the internal shock model.
However, the model would be severely challenged if the upper 
limit continues to go down in the next few years. 

On the other hand, the origin of GRB prompt emission (peaking in the
MeV range) is still not identified. Observations from Swift and Fermi 
observatories suggest that prompt emission is originated from a 
site ``internal'' to the external shock radius \cite{zhang06,abdo09}. 
Among the internal models, besides the internal shock model, other widely 
discussed models include
dissipative photosphere models \cite{rees05,beloborodov10,giannios08} 
and magnetic
dissipation models at large radii \cite{zhang11,lyutikov03}. Recent GRB 
observations with Swift and Fermi missions have challenged the simple
fireball internal shock model \cite{kumar08}, 
and these other mechanisms for GRB prompt 
emission become more attractive. The neutrino signal predictions of these 
prompt emission models could be very different from what is predicted
for the internal shock model. The progressively stringent upper limit
of neutrino flux would start to constrain the validy of these models. 
In this paper, we develop a
general method for calculating the neutrino flux for a wide variety of 
GRB prompt emission models, and discuss how the current upper limit 
constrains these models.

{\em II. General formalism.} 
Our general formalism closely follows the notations adopted by the
IceCube Collaboration \cite{icecube1}, but we make the following changes:
(1) In most previous GRB neutrino flux calculations, the internal 
shock model has been implicitly assumed, so that the radius where
protons are accelerated and the radius where $\gamma$-ray photons 
are generated are both taken as 
\begin{equation}
 R = R_{\rm IS} = \Gamma^2 c \delta t_{\rm min}/(1+z),
\label{RIS}
\end{equation}
where $\delta t_{\rm min}$ is the minimum variability time scale observed 
in a GRB light curve. This widely used expression is valid only for 
internal shocks in a conical jet with the jet opening angle larger
than the relativistic beaming angle $1/\Gamma$, but is not
relevant for most other models. For instance,
in the dissipative photosphere models, the photosphere radius $R_{\rm ph} 
< R_{\rm IS}$, and $\delta t_{\rm min}$ could reflect the intrinsic variability 
time scale of the central engine, which could be larger than the angular
spreading time defined by $R_{\rm ph}/(\Gamma^2 c)$. The large-radius magnetic 
dissipation models (e.g. the Internal Collision-induced MAgnetic
Reconnection and Turbulence or ICMART model \cite{zhang11}) 
can have a GRB emission site $R > R_{\rm IS}$. The rapid 
variability time scale $\delta t$ in these models is related to
the time scale of relativistic mini-jets in the emission region driven by 
relativistic turbulence or reconnection \cite{lyutikov03,narayan09,zhang11}. 
To account for these possibilities, in our formalism we consider the primary
parameters to be $R$ and $\Gamma$ instead of $\delta t$ and $\Gamma$
(see also \cite{murase,he12}).
(2) In the internal shock model, $\gamma$-rays and neutrinos are generated
by electrons and protons accelerated by the same shocks. A parameter
$f_e$ (non-thermal electron-to-proton energy ratio in the internal shocks,
which for $p=2$ takes a value $\sim 0.1$ if the GRB cosmic ray flux 
is normalized to the UHECR flux \cite{waxman95})  
relates the neutrino flux to the observed $\gamma$-ray flux.
In the general formalism, we allow $\gamma$-ray photon production 
and proton acceleration to occur in different locations. 
We therefore introduce a more general parameter $f_{\gamma/p}$
(Eq.\ref{fgamma/p}) to denote the ratio between photon luminosity 
and non-thermal proton luminosity, which reduces to $f_e$ in any model 
that invokes a same site for photon production by leptons and 
proton acceleration (e.g. the internal shock model).
(3) We generalize the previous low optical-depth treatment to also
include a very high optical-depth regime by invoking a more general
$\epsilon_{\nu,1}$ (Eq.\ref{eps_nu_1}). See also \cite{murase08,gao12}
for treatments in the high optical-depth regime.
(4) We introduce another factor $f_p$ (Eq.\ref{fp}) that represents the 
fraction of energy in those protons that can most efficiently 
produce neutrinos via the photo-pion process \cite{li12}.

The general formalism for calculating neutrino flux is as follows:
For an observed ``Band''-function photon flux spectrum 
\begin{eqnarray}
 F_\gamma(E_\gamma) &  = & \frac{dN(E_\gamma)}{d E_\gamma} \nonumber \\
 & = & f_\gamma  \left\{
 \begin{array}{ll}
  \left(\frac{\epsilon_\gamma}{\rm MeV}\right)^{\alpha_\gamma}
  \left(\frac{E_\gamma}{\rm MeV}\right)^{-\alpha_\gamma}, & 
E_\gamma < \epsilon_\gamma \nonumber \\
  \left(\frac{\epsilon_\gamma}{\rm MeV}\right)^{\beta_\gamma}
  \left(\frac{E_\gamma}{\rm MeV}\right)^{-\beta_\gamma}, & 
E_\gamma \geq \epsilon_\gamma
 \end{array},
\right.
\end{eqnarray}
the observed neutrino number spectrum can be expressed as \cite{waxman97, icecube1}
\begin{eqnarray}
& F_\nu(E_\nu) = \frac{dN(E_\nu)}{d E_\nu} 
\nonumber \\
=f_\nu & \left\{
 \begin{array}{ll}
  \left(\frac{\epsilon_{\nu,1}}{\rm GeV}\right)^{\alpha_\nu}
  \left(\frac{E_\nu}{\rm GeV}\right)^{-\alpha_\nu}, & 
E_\nu < \epsilon_{\nu,1} \nonumber \\
  \left(\frac{\epsilon_{\nu,1}}{\rm GeV}\right)^{\beta_\nu}
  \left(\frac{E_\nu}{\rm GeV}\right)^{-\beta_\nu}, & 
\epsilon_{\nu,1} \leq E_\nu < \epsilon_{\nu,2} \nonumber \\ 
  \left(\frac{\epsilon_{\nu,1}}{\rm GeV}\right)^{\beta_\nu}
  \left(\frac{\epsilon_{\nu,2}}{\rm GeV}\right)^{\gamma_\nu-\beta_\nu}
  \left(\frac{E_\nu}{\rm GeV}\right)^{-\gamma_\nu}, & 
 E_\nu \geq \epsilon_{\nu,2}
 \end{array},
\right.
\end{eqnarray}
where
\begin{equation}
 \alpha_\nu = p+1-\beta_\gamma, ~\beta_\nu = p+1-\alpha_\gamma,
~\gamma_\nu = \beta_\nu + 2,
\end{equation}
and $p$ is the proton spectral index defined by $N(E_p)dE_p
\propto E_{p}^{-p} dE_p$. 
The indices $\alpha_\nu$ and $\beta_\nu$ are derived by assuming
that the neutrino flux is proportional to the $p\gamma$ optical depth
$\tau_{p\gamma}$. This is when the fraction of proton energy
that goes to pion production, i.e.
$f\equiv 1-(1-<\chi_{p\rightarrow \pi}>)^{\tau_{p\gamma}}$, is proportional
to $\tau_{p\gamma}$
($<\chi_{p\rightarrow\pi}> \simeq 0.2$ is the average fraction of
energy transferred from protons to pions), which is roughly valid 
when $\tau_{p\gamma} < 3$. In this case, the first break 
\begin{equation}
 \epsilon_{\nu,1} = \epsilon_{\nu,1}^0 \equiv 
7.3\times 10^5 ~{\rm GeV}~(1+z)^{-2}~
\Gamma_{2.5}^2 \epsilon_{\rm \gamma,MeV}^{-1}
\end{equation}
is defined by the break in the photon spectrum. For $\tau_{p\gamma} > 3$,
the $f$ parameter exceeds $\sim 50\%$ and quickly approaches $100\%$. 
The neutrino flux no longer 
significantly increases with $\tau_{p\gamma}$.  If the ``peak''
$p\gamma$ optical depth (the one for protons with energy $E_{p}^p$ to
interact with photons at the peak energy $E_\gamma^p\equiv\epsilon_\gamma$)
\begin{equation}
 \tau_{p\gamma}^p \equiv \tau_{p\gamma}(E_{p}^p)
\simeq \frac{\Delta R'}{\lambda'_{p\gamma}(E_{p}^p)}=0.8 L_{\gamma,52}
\Gamma_{2.5}^{-2} R_{14}^{-1} \epsilon_{\rm \gamma,MeV}^{-1}
\label{fpi}
\end{equation}
is larger than 3, the neutrino
spectrum may be still approximately delineated as the broken 
power law form, but $\epsilon_{\nu,1}$ is smaller by a factor
$(\tau_{p\gamma}^p/3)^{\beta_\gamma-1}$. In general, one can write
\begin{equation}
 \epsilon_{\nu,1} = \epsilon_{\nu,1}^0 {\rm min}
(1,(\tau_{p\gamma}^p/3)^{1-\beta_\gamma}).
\label{eps_nu_1}
\end{equation}
Here $\lambda'_{p\gamma}(E_{p}^p)$ is the comoving proton mean free path 
for $p\gamma$ interaction at $E_{p}^p$, and $\Delta R'$ is the 
comoving width of the jet.
The parameter $R$ denotes the distance of proton acceleration
site (rather than the photon emission site if the two sites are
different) from the central engine. 
The second break energy in the neutrino spectrum,
\begin{equation}
 \epsilon_{\nu,2} = 3.4\times 10^8~{\rm GeV}~ (1+z)^{-1}~
\epsilon_{_B}^{-1/2} L_{w,52}^{-1/2} \Gamma_{2.5}^2 R_{14}
\label{eps2}
\end{equation}
is defined by the $\pi^+$ synchrotron cooling effect, above which 
the newly produced $\pi^+$ lose energy in a time scale shorter 
than the pion decay time scale. Here $\epsilon_{_B}$ is the fraction
of dissipated jet energy in magnetic fields, and $L_w$ is the luminosity 
of the dissipated wind. We further define
\begin{equation}
 f_{\gamma/p}\equiv \frac{L_\gamma}{L_p}, 
\label{fgamma/p}
\end{equation}
and
\begin{eqnarray}
 f_{p} & \equiv & \frac{\int_{E_{p,1}}^{E_{p,2}} dE_p E_p^2 dN(E_p)/dE_p}
{\int_{E_{p,min}}^{E_{p,max}}dE_p E_p^2 dN(E_p)/dE_p} \nonumber \\
& \simeq & \frac{\ln (\epsilon_{\nu,2}/\epsilon_{\nu,1})}
{\ln (E_{p,max}/E_{p,min})}~({\rm for}~p=2),
\label{fp}
\end{eqnarray}
where $E_{p,1}$ \& $E_{p,2}$ 
are proton energies corresponding to $\epsilon_{\nu,1}$ and 
$\epsilon_{\nu,2}$, respectively (Eq.\ref{EnuEp}),
and $E_{p,max}$ and $E_{p,min}$ are the maximum and minimum proton
energy. One can then normalize the neutrino spectrum with the total 
photon fluence, i.e.
\begin{eqnarray}
 \int_0^\infty dE_\nu E_\nu F_\nu(E_\nu) & = &
 \frac{1}{8} \frac{f_p}{f_{\gamma/p}} 
[1-(1-<\chi_{p\rightarrow \pi}>)^{\tau_{p\gamma}^p}] \nonumber \\
& \times & \int_{\rm 1~keV}^{\rm 10 MeV}
dE_\gamma E_\gamma F_\gamma(E_\gamma).
\end{eqnarray}
The coefficient 1/8 is the product of 1/4 (4 leptons share the energy of
one $\pi^+$) and 1/2 (on average roughly half of $p\gamma$ interactions
go to the $\pi^+$ channel when all the $\pi^+$ processes besides
$\Delta^+$ resonance, e.g. direct-pion production, and multiple pion
production, are taken into account).

{\em III. Model-dependent neutrino flux.} Below we apply the general 
formalism to different models. 

(1) Internal shock (IS) model: For typical values of $\delta t_{\rm min}$
and $\Gamma$ as observed or constrained from data, both
photon emission and proton acceleration occur at a typical
internal shock radius $R_{\rm IS} \sim 10^{13}-10^{14}$ cm 
(Eq.\ref{RIS})\cite{footnote}. 
One can simplify the formalism by taking 
$f_{\gamma/p}=f_e$ and $L_w=L_{\gamma}/\epsilon_e$, where 
$\epsilon_e$ is the fraction of the dissipated jet energy in 
electrons (fast cooling assumed). Our formalism is then reduced to
the IceCube formalism \cite{icecube1}, except the additional $f_p$
correction factor and the modification of $\epsilon_{\nu,1}$
(Eq.\ref{eps_nu_1}).

(2) Dissipative photosphere (ph) model: According to this model, the
prompt GRB spectrum is formed near the Thomson scattering photosphere
$R_{ph} \simeq 3.7\times 10^{11} ~{\rm cm}~L_{w,52} \Gamma_{2.5}^{-3}$
\cite{meszaros00}.
In order to account for the observed non-thermal photon spectrum,
it is required that significant energy dissipation and particle
acceleration occur at moderate optical depths
\cite{rees05,beloborodov10,giannios08}. The heating processes
include small-radius internal shocks (those with very short 
variability time scales $\delta t \ll \delta t_{min}$, so that
the internal shock radius is smaller than $R_{ph}$), neutron-proton
collisional heating, or magnetic dissipation in a jet with a 
``striped wind'' magnetic field configuration. 
Efficient proton acceleration is likely in these shocks or magnetic
dissipation site. 
For a same $f_{\gamma/p}$ value, the photosphere model predicts 
a larger $\tau_{p\gamma}^p$ than the IS model by a factor of 
$R_{\rm IS}/R_{ph}$, so that neutrino production is enhanced.

Two mechanisms may lower the neutrino flux for the dissipative 
photosphere models. One is that $f_{\gamma/p}$ could be higher
than 0.1 in a dissipative photosphere. This is especially relevant
for a radiatively efficient photosphere in a high entropy fireball.
The second possibility is that proton acceleration is inefficient
near a dissipative photosphere so that protons are not accelerated
to high enough energies to satisfy the requirement of Eq.\ref{pgamma} 
for pion production. This second possibility deserves more studies,
but the known particle-in-cell simulations favor acceleration
of protons in mildly relativistic internal shocks 
even if the magnetization parameter $\sigma$ (the ratio between
the Poynting flux and the matter kinetic flux) 
is as high as 0.1 \cite{sironi09}, especially in a 
striped-wind magnetic configuration \cite{sironi11}.

(3) Photosphere + internal shock (ph+IS) model: For any efficient
dissipative photosphere model, $\sigma \leq 1$ is expected at the 
photosphere (otherwise the photosphere luminosity
is suppressed by a factor $(1+\sigma)$).
Internal shocks would in any case develop at $R_{\rm IS} > R_{ph}$
($\sigma \ll 1$), where protons are accelerated. Even if 
photon emission at internal shocks may be inefficient, photons 
emitted from the photosphere would in any case
pass through the internal shock region
and interact with the energetic protons there to produce neutrinos.
Due to dissipation, on average, the Lorentz factor in the internal
shocks is expected to be somewhat smaller than that in the photosphere.
The comoving photon number density in the internal shock region 
would be somewhat higher at $R_{\rm IS}$ in the ph+IS model than in
the IS model. This tends to increase the neutrino flux with respect
to the IS model. The photon flux received by the IS protons is 
anisotropic. However, for a roughly isotropic distribution of protons
in the comoving frame as commonly envisaged, the neutrino production 
efficiency would not be significantly
reduced \cite{fan05}. Finally, the parameter $f_{\gamma/p}$ 
can be larger than $f_e$ due to the efficient photon production in
the photosphere than in internal shocks.
Considering all these factors, we expect that the predicted neutrino
flux level in the ph+IS model is roughly the same (within a factor of
a few) as that in the IS model.

(4) The ICMART and other large-radius magnetic dissipation 
models: The ICMART model \cite{zhang11} invokes a highly magnetized 
outflow, which remains un-dissipated up to a radius $R_{\rm ICMART} >
R_{\rm IS}$. Emission from the photosphere and internal shocks is
greatly suppressed. Internal shocks help to destroy the ordered magnetic 
fields, and a strong run-away magnetic dissipation process occurs at a 
large radius $R_{\rm ICMART} \sim \Gamma^2 \delta t_{\rm slow}
\sim 10^{15}$ cm \cite{zhang11}, where $\delta t_{\rm slow} \gtrsim 1$s is
the slow variability component in the GRB lightcurves \cite{gao12b}.
The $\tau_{p\gamma}^p$ parameter is smaller by a factor 
$R_{\rm ICMART}/R_{\rm IS}\sim (10-100)$ than the IS model. 
This model therefore predicts a much lower neutrino flux than the IS 
model for a same $f_{\gamma/p}$ value.

(5) External shocks: even though MeV $\gamma$-rays are produced
at an internal radius \cite{zhang06}, these photons must pass through
the external shock region. Due to the large radius of the external shock,
the optical depth of $p\gamma$ interaction is much lower, and the
neutrino production is much less efficient than the other processes
discussed above.

\begin{figure}[t!]
\vspace{0.3cm}
{\centering
\resizebox*{0.5\textwidth}{0.3\textheight}
{\includegraphics{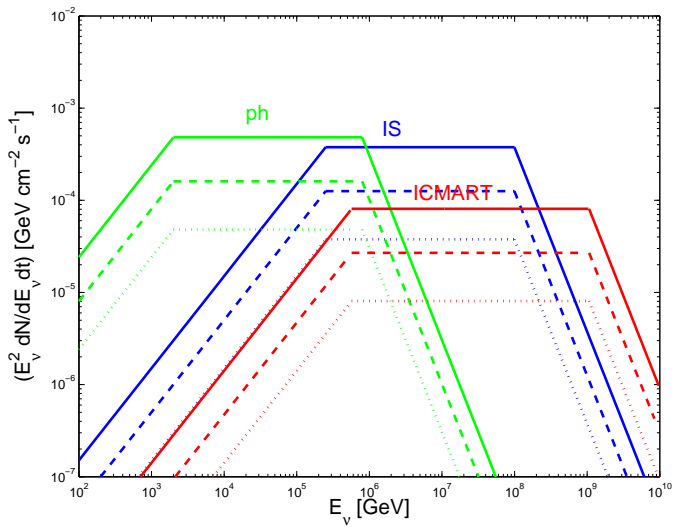}}
\par}
\caption{The predicted neutrino flux for a typical GRB in three
GRB prompt emission models: ``ph'' (green): dissipative photosphere model;
``IS'' (blue): internal shock model; ``ICMART'' (red): internal-collision-induced
magnetic reconnection and turbulence model. Model parameters:
$L_{\gamma,52}=1$, $\delta t=0.1$ s, $\epsilon_{\gamma,\rm MeV}
=0.2$, $\alpha_\gamma=1$, $\beta_\gamma=2$, $p=2$, $z=1$, 
$\Gamma=250$, $\epsilon_{_B}/\epsilon_e = 1$, $R_{\rm ICMART}=10^{15}$ cm. 
Three values of $f_{\gamma/p}$ are adopted: 0.1 (solid), 0.3 (dashed),
and 1 (dotted).}
\label{spectrum}
\end{figure}

We calculate the neutrino flux of a typical GRB in different models
in Fig.1. The following values for the measurable parameters are adopted:
$L_{\gamma,52}=1$, $\delta t_{\rm min}=0.1$ s, $\epsilon_{\gamma,\rm MeV}
=0.2$, $\alpha_\gamma=1$, $\beta_\gamma=2$, $p=2$, and $z=1$
\cite{p}. We plot
three models: the photosphere model (`ph', green), the internal shock model
(`IS', blue), and the ICMART model (`ICMART', red). Since these are all
one-zone models, we only have three
free parameters: $\Gamma$, $f_{\gamma/p}$ and $\epsilon_{_B}/\epsilon_e$.
We take a conventional value $\epsilon_{_B}/\epsilon_e \sim 1$ in our 
calculation of $\epsilon_{\nu,2}$. Since the dependence is shallow (1/2 power),
a more precise treatment of the ratio based on a fundamental understanding
of particle acceleration physics would not significantly alter the results.

The predicted neutrino flux is sensitive to $\Gamma$
(e.g. $\tau_{p\gamma}^p \propto \Gamma^{-4}$ in the IS model). Instead 
of using the ``benchmark'' value $\Gamma=300$ \cite{icecube1,icecube2},
we use the values inferred from various observational constraints
\cite{sari99,liang10,lv12},
which led to a correlation between $\Gamma$ and isotropic luminosity
\cite{liang10,lv12}:
\begin{equation}
 \Gamma \simeq 250 L_{\gamma,52}^{0.30}.
\label{Gamma-L}
\end{equation}
This gives $\Gamma \sim 250$ for the example GRB, which gives a stronger
neutrino flux due to the strong $\Gamma$-dependence on the neutrino flux.

If GRBs are the dominant UHECR sources, than the proton flux from GRBs
can be normalized by the observed UHECR sources, which requires 
$f_{\gamma/p} = 0.1$ for $p=2$ \cite{waxman95,waxman97}
(solid lines in Fig.1). Since some models (e.g. dissipative photosphere
models and magnetic dissipation models) can have a higher $f_{\gamma/p}$ value,
which can interpret the GRB data well without requiring GRBs as the 
dominant sources of UHECRs, we also plot the flux levels 
of the three models for two other larger values of $f_{\gamma/p}$: 
0.3 (dashed lines) and 1 (dotted line).

{\em IV. Current status and future prospects.} 
The continued search for neutrino signals from GRBs by the IceCube 
Collaboration is starting to pose meaningful constraints on GRB models.
With the current limit, the IS model with $f_{\gamma/p}=0.1$ and $\Gamma-L$ correlation
just starts to barely violate the observational constraint \cite{he12}.
For the same value of $f_{\gamma/p}=0.1$, the dissipative photosphere (ph) 
models are already disfavored, unless an unknown mechanism suppresses
 proton acceleration in the photosphere region. On the other hand,
an radiative efficient dissipative photosphere model may allow $f_{\gamma/p}$ 
to be larger. These models may be constrained with even deeper upper
limits  (Fig.1, see also \cite{gao12}).
The ICMART model and other large-scale
magnetic dissipation models are entirely consistent with the data.
Thanks to the low neutrino background in the interested energy range
in coincidence with GRBs in time and direction, 
the upper limit would go down linearly with time. In a few more years,
if high energy neutrinos are still not detected from GRBs, one would 
either require a large $f_{\gamma/p}$, or demand a larger emission 
radius than the internal shock radius, as expected in some magnetic
dissipation models such as the ICMART model.

We thank Zhuo Li, Peter M\'esz\'aros, Kohta Murase, 
Soebur Razzaque, and Xiang-Yu Wang for useful discussion,
the referees for helpful comments.
This work is partially supported by NSF through
AST-0908362 (BZ) and AST-0909110 (PK). BZ acknowledges the Cheung Kong
Scholar fellowship and the Kavli Institute for Astronomy and Astrophysics,
Peking University for support during his sabbatical leave from UNLV.


\begin{thebibliography}{55}
\bibitem{footnote}
A larger IS radius may be obtained if the
$\delta t_{\rm min}$ is taken not as the central engine variability
time scale, as invoked in some more complicated two-zone models.
\bibitem{p}
For easy comparison among different models, a universal $p=2$ is
adopted for all the models. In reality, different $p$ values may result
from different acceleration mechanisms.
\end{thebibliography}
\end{document}